\documentstyle[aps,prl,preprint]{revtex}
\begin{document}
\draft
\preprint{\small\rm Universit\`a di Catania,
Dipartimento di Fisica, preprint 11/95}
\title{Are light alkali metals still metals under high pressure?}
\author{Fabio Siringo, Renato Pucci and Giuseppe G.N. Angilella}
\address{Dipartimento di Fisica dell'Universit\`a di Catania,
57, Corso Italia, I-95129 Catania, Italy}
\date{\today}
\maketitle
\begin{abstract}
The extended Hubbard Hamiltonian on a bcc lattice is studied at half-filling
and for a finite hopping between next-nearest-neighbours, 
in mean-field 
approximation. An ionic insulating broken-symmetry phase is predicted for any
hydrogenoid bcc solid in the density range $1.0<r_s <2.6$. The occurrence
of an ionic phase would explain the failure to achieve hydrogen metallization
at high pressures.
Moreover, a metal-insulator transition is expected for sodium in the
$100$~GPa region.
\end{abstract}

Alkali metals crystallize in a bcc phase under ordinary thermodynamic
conditions. 
In 1935, Wigner and Huntington~\cite{Wigner} proposed that even
hydrogen should undergo a metal-insulator (MI) transition from a molecular
to a monatomic bcc crystal, under high pressure in analogy to alkali
metals. During the last years
new excitement arose after several claims for the reach of hydrogen's
metallization~\cite{Eggert,Mao,Cui,Hanfland}.
However, it is now out of doubt
that all the observed phases of hydrogen are still molecular, and 
there is no evidence of metallization induced by a band overlap
mechanism up to 191~GPa~\cite{Chen}. From this point of view,
hydrogen seems to be quite different from the halogens,
for which MI transitions have been 
observed~\cite{Pucci,Fujihisa}.

The nature of the high pressure $A$ phase of hydrogen still remains
unexplained, and several hypotheses have been advanced. The recent
proposal by Baranowski~\cite{Baranowski} that the hydrogen molecules
may develop electric dipole moments
is challenging for different reasons:
\emph{i)} first of all, such a phase would be \emph{per se} interesting,
being a broken-symmetry ground state of the symmetric H$_2$ molecular
system; \emph{ii)} in addition, the existence of such an ionic phase
could move a possible monatomic phase further towards higher pressures;
\emph{iii)} finally, given the similarity of hydrogen with the alkali
metals, the possible existence of an ionic instability should show up
even in lithium or sodium under the proper thermodynamic conditions.

Provided that such a broken-symmetry ionic crystal does exist, it would
be desirable to fix the boundaries between the ionic and the monatomic 
bcc phases.
Approaching the boundary from the bcc phase, we look for an ionic 
instability of the monatomic crystal, or, in other words, for a
charge density wave (CDW) instability, commensurate with the cubic lattice.
The existence of such a CDW has been recently observed in the ground state of 
sodium and potassium, under ordinary thermodynamic 
conditions~\cite{Giebultowicz}. Such instabilities of the Fermi gas had
been predicted~\cite{Overhauser} as a consequence of the Coulomb
electron-electron ($e$-$e$) 
repulsion, but do not give rise to any MI transistion,
since the CDW is not commensurate with the lattice. Such very small effect
does not prevent us from considering the alkali metals as simple `free-electron'
metals for most aspects. On the other hand, a nearest-neighbour tight-binding
model on a bipartite bcc lattice gives rise to a perfectly nested cubic Fermi
surface at half-filling, and any small $e$-$e$ repulsion would drive
the system towards a spin density wave (SDW) or towards a CDW commensurate
with the lattice.


In this Letter we show that, even without nesting of the Fermi surface,
an hydrogenoid bcc crystal undergoes a MI transition towards a 
broken-symmetry commensurate CDW phase, for an appropriate bounded 
range of density values. Such a conclusion emerges from a careful
analysis of the mean-field phase diagram for an extended Hubbard Hamiltonian,
modified in order to take in due account  the long range Coulomb interactions
and the hopping between next-nearest-neighbours. Even for a spherical Fermi
surface, the model predicts the occurrence of a broken-symmetry insulating
ground state, provided that the nearest-neighbour repulsive interaction
$V$ exceeds some critical value. The latter is a function of the other
energy scales and mainly of the on-site Hubbard repulsion $U$ between
two electrons sharing the same lattice site. While $U$ is only slightly
affected by any increase in density, $V$ scales as $a^{-1}$, being
$a$ the cubic lattice spacing. Under high pressure, $V$ may reach its
critical value, giving rise to a MI transition, albeit in an intermediate
density range; then, at very high densities, the large increase of the Fermi
energy, scaling as $a^{-2}$, would eventually stabilize the monatomic
phase. In other words, we expect that under high pressure both lithium
and sodium should undergo a MI transition from a simple metal to an ionic
insulator. On the other hand, such a high density instability of
the bcc crystal would suggest that, in order to stabilize a hydrogen
monatomic phase, higher densities are required than previously estimated.

The extended Hubbard Hamiltonian reads as:
\begin{eqnarray}
H && = -t_1 \sum_{\langle ij\rangle\, \sigma} c^{\dag}_{i\sigma} c_{j\sigma}
-t_2 \sum_{\langle\langle ij\rangle\rangle\,\sigma}
c^{\dag}_{i\sigma} c_{j\sigma} \nonumber \\ 
 && + U\sum_i n_{i\uparrow} n_{i\downarrow}
+ \frac{1}{2} \sum_{i\neq j\,\sigma\sigma^\prime} V_{ij} 
n_{i\sigma} n_{j\sigma^\prime} ,
\label{eq:Hubbard}
\end{eqnarray}
where $c_{i\sigma}$ ($c_{i\sigma}^{\dag}$) denote the annihilation (creation) 
operators for an electron in the Wannier state centered on the $i$th site
of a bcc lattice, with spin projection $\sigma\in\{\uparrow,\downarrow\}$
$n_{i\sigma} = c_{i\sigma}^{\dag} c_{i\sigma}$, $t_1$, $t_2 > 0$,
and $\langle ij\rangle$, $\langle\langle ij\rangle\rangle$ restricting the
sums over nearest and next-nearest-neighbour couples, respectively.

The Hamiltonian, Eq.~(\ref{eq:Hubbard}), incorporates two major 
approximations:
\emph{i)} it neglects all bond-bond and bond-ion interactions; \emph{ii)} 
it neglects any hopping term other than those between nearest or 
next-nearest
neighbours. 
Approximating the Wannier states by atomic hydrogen ground
state wave functions, for a density corresponding to $r_s \approx 2.2$,
the larger bond-ion interaction term does not exceed the 30\% of the
corresponding ion-ion interaction. Of course, any extrapolation to higher 
densities would require some caution. Regarding the neglected hopping terms,
we must notice that the insertion of such exponentially decreasing terms
does not change the shape of the Fermi surface in a significant way.
A ratio $t_2 /t_1 \approx 0.6\div 0.7$ allows for an almost spherical
Fermi surface, up to a 3\% deviation.

The model may be solved in mean-field (MF) approximation by inserting
$\langle n_{i\sigma} \rangle = \frac{1}{2} + \Delta_{i\sigma}$ and
neglecting second order terms in the fluctuations $\delta n_{i\sigma}
= n_{i\sigma} - \langle n_{i\sigma} \rangle$. Since we are looking
for a commensurate CDW instability, we assume
$\Delta_{i\sigma} = \Delta\cos ({\mathbf Q}\cdot{\mathbf R}_i )$,
being ${\mathbf Q} = \left( \frac{2\pi}{a} ,0,0 \right)$ the exact
nesting vector of the Fermi surface when $t_2 =0$. Working in the reciprocal
lattice, \emph{i.e.} introducing:
\begin{equation}
c_{i\sigma}^{\dag} = \frac{1}{\sqrt{N}} \sum_{{\mathbf k}}
{\mathrm e}^{-i{\mathbf k}\cdot{\mathbf R}_i} c_{{\mathbf k}\sigma}^{\dag} ,
\end{equation}
with ${\mathbf k}$ summed over the $N$ points inside the first Brillouin
zone, and neglecting second order fluctuation terms, the Hamiltonian
Eq.~(\ref{eq:Hubbard}) reads (up to a constant) as:
\begin{equation}
H_{\mathrm MF} = \sum_{{\mathbf k}\sigma} \varepsilon({\mathbf k})
c_{{\mathbf k}\sigma}^{\dag} c_{{\mathbf k}\sigma}
+ \Gamma \sum_{{\mathbf k}\sigma}
c^{\dag}_{{\mathbf k}+{\mathbf Q}\sigma} c_{{\mathbf k}\sigma} 
  -N\Gamma\Delta ,
\label{eq:HubbardMF}
\end{equation}
where $\Gamma=\Delta (U-16W)$, $\varepsilon({\mathbf k})=
\varepsilon_1 ({\mathbf k}) + \varepsilon_2 ({\mathbf k})$, and:
\begin{eqnarray}
\label{eq:dispersion1}
\varepsilon_1 ({\mathbf k}) && = -4t_1 
\cos\left( \frac{k_x a}{2} \right)
\cos\left( \frac{k_y a}{2} \right)
\cos\left( \frac{k_z a}{2} \right), \\
\label{eq:dispersion2}
\varepsilon_2 ({\mathbf k}) && = -t_2 
[\cos (k_x a) + \cos (k_y a) + \cos (k_z a) ].
\end{eqnarray}
Here $W$ is a renormalized interaction parameter summing up all
the long range Coulomb interactions:
\begin{equation}
\label{eq:RenormalizedW}
W = \frac{1}{8} \sum_{m=1}^{\infty} (-1)^{m+1} z_m V^{(m)} ,
\end{equation}
where, for $m=1,2,\ldots\infty$, $V^{(m)}$ denotes the $V_{ij}$
interaction term for nearest-neighbours, next-nearest-neighbours,
\emph{etc.,} and $z_m$ is the corresponding coordination number.

The MF Hamiltonian, Eq.~(\ref{eq:HubbardMF}), is easily diagonalized
by a canonical transformation. Let us introduce the spinorial notation
$\Psi^{\dag}_{{\mathbf k}\sigma} = \left(
c_{{\mathbf k}\sigma}^{\dag} ~,~~ c_{{\mathbf k}+{\mathbf Q}\sigma}^{\dag}
\right)$,
with ${\mathbf k}$ restricted inside the cube $|k_\alpha | < \pi /a$
($\alpha=x,y,z$), which is exactly half the first Brillouin zone.
Actually the transformation ${\mathbf k} \rightarrow {\mathbf k}+{\mathbf Q}$
maps such reduced zone onto the complementary half-zone. The Hamiltonian
Eq.~(\ref{eq:HubbardMF}) now reads:
\begin{equation}
H_{\mathrm MF} = \sum_{{\mathbf k}\sigma} \Psi^{\dag}_{{\mathbf k}\sigma}
h({\mathbf k} ) \Psi_{{\mathbf k}\sigma} -N\Gamma\Delta,
\end{equation}
where the $2\times 2$ matrix $h({\mathbf k})$ is defined as:
\begin{equation}
h({\mathbf k}) = \left(
\begin{array}{cc}
\varepsilon_1 ({\mathbf k}) + \varepsilon_2 ({\mathbf k}) &
\Delta (U -16 W) \\
\Delta (U -16 W)  & 
\varepsilon_2 ({\mathbf k}) - \varepsilon_1 ({\mathbf k})
\end{array} \right),
\end{equation}
and is promptly diagonalized yielding the spectrum:
\begin{equation}
E^\pm ({\mathbf k}) = \varepsilon_2 ({\mathbf k} ) \pm
\sqrt{\varepsilon_1 ({\mathbf k})^2 + \Gamma^2}.
\end{equation}
A gap opens between the two bands, $E^\pm ({\mathbf k})$,
whenever $\Gamma > 2t_2$: in such a regime, the system is an insulator,
and the total energy $E_{\mathrm tot}$ 
is readily evaluated by summing $E^-$ over the doubly occupied half-zone:
\begin{equation}
\label{eq:TotalEnergy}
E_{\mathrm tot} = \sum_{{\mathbf k}\sigma}
[\varepsilon_2 ({\mathbf k}) - \sqrt{\varepsilon_1 ({\mathbf k})^2 + \Gamma^2}
] - N \Gamma\Delta.
\end{equation}
A gap equation is obtained by differentiating
$E_{\mathrm tot}$ with respect to the order parameter, $\Delta$:
\begin{equation}
\label{eq:GapEquation}
\frac{1}{16W -U} = \frac{a^3}{2} \int 
\frac{{\mathrm d}^3 {\mathbf k}}{(2\pi)^3}
\frac{1}{\sqrt{\varepsilon_1 ({\mathbf k})^2 + \Gamma^2}} .
\end{equation}
A finite $\Gamma$ always solves the latter condition at any coupling
strengths but, for sake of consistency with the above assumption of
dealing with an insulating phase, $\Gamma$ must exceed the critical value,
$\Gamma_c = 2t_2$. A consistent minimum for the total
energy is \emph{e.g.} found for 
$(16W -U)/t_1 >3.84$ if $t_2 /t_1 =0.8$. 

In the insulating phase, the Coulomb interaction 
is not screened by the conduction electrons, and long range contributions
cannot 
be neglected. We should evaluate the parameters $U$, $V^{(m)}$
as diagonal matrix elements of the bare Coulomb interaction in the Wannier 
representation.
If we assume
$V_{ij} \sim 1/ |{\mathbf R}_i - {\mathbf R}_j |$,
the renormalized parameter $W$ follows from its 
definition, Eq.~(\ref{eq:RenormalizedW}), as
$W\approx \frac{1}{8} \alpha_{\mathrm M} V$,
being $\alpha_{\mathrm M}$ the Cs--Cl Madelung constant,
$\alpha_{\mathrm M} = 1.763$~\cite{Madelung}, and $V\equiv V^{(1)}$
the first term in the expansion, Eq.~(\ref{eq:RenormalizedW}), 
\emph{i.e.,} the nearest-neighbour repulsive interaction. The problem
is thus mapped back onto the standard extended Hubbard model with an effective
number of nearest-neighbour sites $z^* = \alpha_{\mathrm M}$, to be
compared with the bcc value $z_1 =8$. 
For the reasonable choice $t_2 /t_1 =0.8$~\cite{comment3},
the $U$--$V$ phase diagram for the CDW instability is shown in 
Fig.~\ref{fig:PhaseDiagram}.

The boundary between the metallic and the ionic insulating phases is
given by the simple linear relation 
$U/t_1 = -\gamma + 2\alpha_{\mathrm M} V/t_1$,
being $\gamma$ 
the minimum value of the ratio $(16W-U)/t_1$, as emerging from
Eq.~(\ref{eq:GapEquation}) for $\Gamma=\Gamma_c =2t_2$
($\gamma=3.84$ for $t_2 /t_1 =0.8$). In principle, an ionic metallic
phase may exist just over the boundary, since the gap closure is due
to band overlap, while a finite order parameter $\Delta$ always arises
from the gap equation~(\ref{eq:GapEquation}). However, in presence of a band
overlap, the total energy, Eq.~(\ref{eq:TotalEnergy}), is incorrect, since
the energy levels must be summed up to the Fermi value inside 
both the bands $E^\pm$.
Thus, Eq.~(\ref{eq:GapEquation}) is not correct in the metallic phase
and the existence of a stable broken-symmetry ground state is questionable
in the metallic regime. Moreover, all the Coulomb interaction terms would be
strongly screened by the conduction electrons, so that the symmetric $\Delta
=0$ ground state is expected to be more favoured for the metallic phase.
The phase diagram is incomplete, since we have not taken in consideration
the possible occurrence of 
SDW instabilities, which are likely to be present for 
$U\gg V$, though irrelevant for the following considerations.

Let us first discuss the $U$--$V$ phase diagram in relation with the
behaviour of solid hydrogen under pressure. The possible bcc phase
of dense hydrogen would be a simple `free-electron' metal, with an almost
spherical Fermi surface and a Fermi energy comparable with the free
electron value $E_{\mathrm F} = 1.84 /r_s^2$~a.u., which also reproduces the
observed Fermi energies for the alkali metals. In the metallic phase,
neglecting any interaction term, the model Fermi energy arises from the
unperturbed spectrum given by Eqs.~(\ref{eq:dispersion1}) and (\ref{eq:dispersion2})
in terms of the parameters $t_1$, $t_2$ as 
$E_{\mathrm F} = \varepsilon ({\mathbf k}_{\mathrm F} )
-  \varepsilon(0)$,
where ${\mathbf k}_{\mathrm F}$ is the Fermi vector, $ak_{\mathrm F}
\approx (6\pi^2 )^{1/3}$. Comparing the latter with the definition for
$E_{\mathrm F}$ in the free-electron case yields an estimate of $t_1$ for
any chosen ratio $t_2 /t_1$, and for any fixed density. 
If we fix $t_2 /t_1 =0.8$, then
$t_1 = 0.259 /r_s^2$~a.u.
Approximating the Wannier states by $1s$ hydrogen wave functions, both $U$ 
and $V$ follow, in a.u.~\cite{Slater}, as
$U = \frac{5}{8}$,
$V = \frac{1}{R} - {\mathrm e}^{-2R}
\left( \frac{1}{R} + \frac{11}{8} + \frac{3}{4} R + \frac{1}{6} R^2 \right)$,
where the nearest-neighbour distance $R= (\sqrt{3} \pi)^{1/3} r_s$ has
been used. For large $r_s$ values, $U/t_1 \sim r_s^2$,
while $V/t_1 \sim r_s$, so that the equation of state for hydrogen
in the $U/t_1$--$V/t_1$ phase diagram of Fig.~\ref{fig:PhaseDiagram}
is just a parabola. Some possible states of dense hydrogen are
reported on Fig.~\ref{fig:PhaseDiagram} 
At very high density, the equation of state
deviates from the parabolic behaviour, 
since $V$ saturates for $r_s \rightarrow 0$.
However, the very high density limit is questionable, and must be considered 
as an extrapolation out of the range 
where the adopted approximations are reasonable.

The phase diagram is not significatively altered by a change of the ratio $t_2 /t_1$, since
both the boundary line and the equation of state are shifted in the same direction
and their relative changes compensate.

If we rely on the emerging scenario, even in the very high density limit,
then we would predict that hydrogen metallization requires $r_s <1$,
since the bcc phase would be unstable towards an insulating ionic phase for
$1.0 <r_s <2.6$. The occurrence of such an ionic phase could explain
the failure of all the attempts to reach the monatomic state at the
currently achievable densities. In fact, as discussed by 
Chen \emph{et al.}~\cite{Chen} 
and Ashcroft~\cite{Ashcroft}, the occurrence of an
IR active vibron mode could be justified by the presence of permanent
dipole moments. Besides, the occurrence of any other molecular ionic
phase cannot be ruled out by our approach, which only prevents the stability
of the monatomic bcc structure for a given high density range. However,
the high density limit $r_s \simeq 1$ is only qualitatively correct, as
previously discussed, being the approximations out of control in this very
high density regime. Therefore, we don't find any contrast with the
Monte Carlo prediction~\cite{Ceperley} of a transition towards the 
monatomic phase for $r_s \approx 1.3$. Moreover, that numerical
calculation evidenced the equivalence of the monatomic ground state energies
in both the fcc and in the bcc phases. Thus the ionic instability,
lowering the ground state energy, should be relevant even if an fcc structure
were the most stable monatomic phase.

While the lower bound of the ionic phase is only qualitatively
determined by the present approach, the upper limit $r_s <2.6$
is much more reliable. Any hydrogenoid bcc solid should undergo a MI 
transition around that critical density value~\cite{comment1}, 
thus realizing an unexpected broken-symmetry ionic phase.
At room pressure, $r_s =3.94$ for sodium and $r_s =3.25$ for
lithium, so that according to Fig.~\ref{fig:PhaseDiagram} both
the elements are correctly predicted to be stable in the monatomic bcc phase.
In order to reach the critical density $r_s \approx 2.6$, a very high pressure
is required~\cite{Olijnyk,Olinger,Alexandrov}.
Such a pressure could be really prohibitive for K, Rb and Cs. Besides,
these heavier alkali elements undergo several pressure induced structural
transitions which are believed to be driven by the electronic transfer
to upper empty bands~\cite{Olijnyk}. Therefore, our attention should be focused on the
lightest alkali, since such transfer mechanism is negligible and the required
pressure could be reached by modern diamond anvil cell apparatus.
Lithium has been compressed up to $r_s \approx 2.8$~\cite{Olinger}, and a
phase transition from bcc to fcc has been observed for $P=6.9$~GPa
($r_s \approx 2.9$). The occurrence of the fcc phase could in principle
invalidate our discussion, even if the ground state energies of such
cubic monatomic structures are so close that the ionic instability
cannot be ruled out at higher densities. 
Mostly, the best alkali
with which we may compare our prediction is sodium: \emph{i)} it has
an almost spherical Fermi surface; \emph{ii)} its $3s$ orbital is expected
to be comparable for extension with the $1s$ orbital of 
hydrogen~\cite{comment2}; \emph{iii)}  no structural phase transition has 
been observed up to $r_s \approx 2.9$~\cite{Alexandrov}; \emph{iv)}
the first empty $d$ band is far from the Fermi energy. A structural phase 
transition has been predicted~\cite{Moriarty} from bcc to hcp 
at the very high density $r_s \lesssim 2.5$, which should
be reachable in the 100~GPa region~\cite{Olijnyk}.
According to Fig.~\ref{fig:PhaseDiagram}, a MI transition
should occur first, around $r_s \approx 2.6$,
then the ionic phase could push to higher densities the 
structural transition.

At this stage, we should reverse our starting question, and we should ask
instead: Why is sodium a metal? In a broken-symmetry ionic phase, the 
inter-ion Coulomb interactions add a considerable contribution
to the total ground state energy. This
very same term amounts to the almost entire cohesive energy of any ionic crystal.
We cannot neglect such interactions, even at high densities, in comparison
with the Fermi energy. On the other hand, the r\^ole played by the on-site
Hubbard $U$ and by the nearest-neighbour effective interaction $W$
is competitive, as it is evident from Eqs.~(\ref{eq:TotalEnergy}) 
and (\ref{eq:GapEquation}). The former favours a SDW instability,
whereas the latter is responsible for the onset of a CDW. In sodium,
under normal thermodynamic conditions, such interaction terms compensate
each other, precluding any instability. The ground state is a metal,
the interactions are strongly screened by the Fermi liquid, and give
only rise to a small renormalization of the band parameters. In other
words, the symmetric metallic phase is based 
on the equilibrium between competing
interactions.
Whenever we alter such an equilibrium (\emph{e.g.} increasing $W$ by
compressing the distances) the system falls into a broken-symmetry
insulating phase, where the interactions are no longer screened
and play an essential r\^ole. 
Such a phase is expected for sodium in the $100$~GPa region, for 
$r_s \approx 2.6$.
The ionic phase could explain the failure to reach hydrogen metallization, and 
would give a possible interpretation of the anomalous IR active vibron
mode observed in the high pressure $A$ phase. 
A very high density is required in order to
reduce the r\^ole of the interactions
in comparison with the kinetic electronic energy,
and to restore symmetry. For a perfectly nested Fermi surface ($t_2 =0$),
such a density is infinite, since the integral in the gap 
equation~(\ref{eq:GapEquation}) diverges for $\Gamma=0$. Out of nesting
($t_2 > 0$), a metallic phase arises around the 
$U\approx 2\alpha_{\mathrm M} V$ 
region (Fig.~\ref{fig:PhaseDiagram}), which also explains the behaviour
of all the alkali metals.

\begin{figure}
\caption{$U$--$V$ phase diagram for a CDW instability, for $t_2 /t_1 =0.8$.
The boundary between normal metal (M) and ionic insulator (I) is reported
as a solid line. The dashed line represents the equation of state for
a light alkali as hydrogen: the squares correspond to $r_s =0.6\div 2.8$.}
\label{fig:PhaseDiagram}
\end{figure}

\begin{references}
\bibitem{comment3} A slight change of the ratio 
$t_2 /t_1$ does not affect too much the phase
diagram. In general, a smaller value than $0.8$ would be expected, thus
giving rise to an enlargement of the ionic region in the phase diagram.
\bibitem{comment1} Notice that for $r_s \sim 2.6$, $V$ is almost independent
of the atomic size, while $U$ is slightly smaller for larger atomic orbitals,
so that the critical $r_s$ value could be even larger for the other alkali
metals.
\end{references}
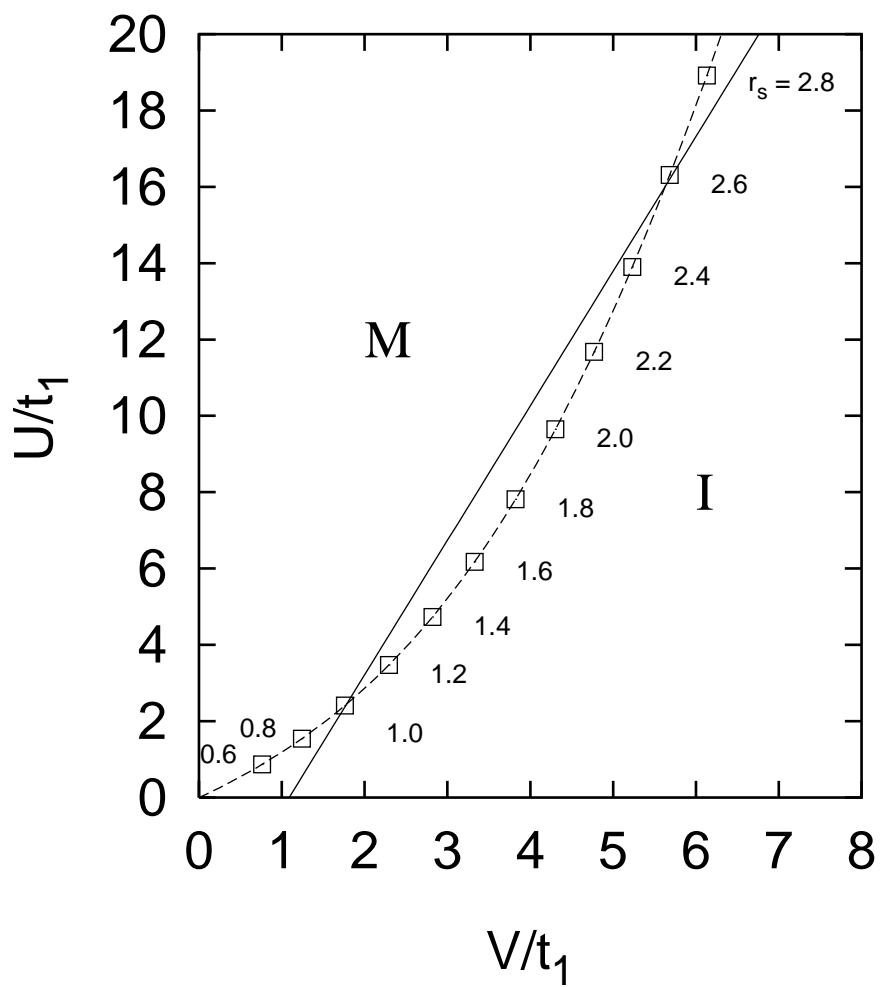
\end{document}